 \documentstyle[aps,prl,12pt,epsf]{revtex}
\def\newblock{\hskip .11em plus .33em minus .07em}
\newcommand{\psik}{{$B^+\rightarrow J/\psi K^+$}}

\newcommand{\pstar}{{$B^{0}\rightarrow J/\psi K^{*0}$}}

\newcommand{\dc}{{$d\sigma/dp_T$}}
\newcommand{\roots}{$\sqrt{s}$}

\begin{document}
\begin{flushright}
FERMILAB-Pub-95/48-E
March 18, 1995
\end{flushright}
\vspace{0.3in}
\begin{center}
\bf
Measurement of the $B$ Meson Differential Cross Section, \\
\dc, in $p\overline{p}$ Collisions at $\sqrt{s} = 1.8$ TeV \\
\end{center}
\vspace{0.3in}
\font\eightit=cmti8
\def\r#1{\ignorespaces $^{#1}$}
\hfilneg
\begin{center}
\begin{sloppypar}
\noindent
F.~Abe,\r {13} M.~G.~Albrow,\r 7 S.~R.~Amendolia,\r {23} D.~Amidei,\r {16}
J.~Antos,\r {28} C.~Anway-Wiese,\r 4 G.~Apollinari,\r {26} H.~Areti,\r 7
M.~Atac,\r 7 P.~Auchincloss,\r {25}
% ...
\end{sloppypar}

\vskip .025in
(CDF Collaboration)
\end{center}

\vskip .025in
\begin{center}
\r 1  {\eightit Argonne National Laboratory, Argonne, Illinois 60439} \\
\r 2  {\eightit Istituto Nazionale di Fisica Nucleare, University of Bologna,
I-40126 Bologna, Italy} \\
\r 3  {\eightit Brandeis University, Waltham, Massachusetts 02254} \\
\r 4  {\eightit University of California at Los Angeles, Los
Angeles, California  90024} \\
\r 5  {\eightit University of Chicago, Chicago, Illinois 60637} \\
\r 6  {\eightit Duke University, Durham, North Carolina  27708} \\
\r 7  {\eightit Fermi National Accelerator Laboratory, Batavia, Illinois
60510} \\
\r 8  {\eightit Laboratori Nazionali di Frascati, Istituto Nazionale di Fisica
               Nucleare, I-00044 Frascati, Italy} \\
\r 9  {\eightit Harvard University, Cambridge, Massachusetts 02138} \\
\r {10} {\eightit University of Illinois, Urbana, Illinois 61801} \\
\r {11} {\eightit Institute of Particle Physics, McGill University, Montreal
H3A 2T8, and University of Toronto,\\ Toronto M5S 1A7, Canada} \\
\r {12} {\eightit The Johns Hopkins University, Baltimore, Maryland 21218} \\
\r {13} {\eightit National Laboratory for High Energy Physics (KEK), Tsukuba,
Ibaraki 305, Japan} \\
\r {14} {\eightit Lawrence Berkeley Laboratory, Berkeley, California 94720} \\
\r {15} {\eightit Massachusetts Institute of Technology, Cambridge,
Massachusetts  02139} \\
\r {16} {\eightit University of Michigan, Ann Arbor, Michigan 48109} \\
\r {17} {\eightit Michigan State University, East Lansing, Michigan  48824} \\
\r {18} {\eightit University of New Mexico, Albuquerque, New Mexico 87131} \\
\r {19} {\eightit Osaka City University, Osaka 588, Japan} \\
\r {20} {\eightit Universita di Padova, Istituto Nazionale di Fisica
          Nucleare, Sezione di Padova, I-35131 Padova, Italy} \\
\r {21} {\eightit University of Pennsylvania, Philadelphia,
        Pennsylvania 19104} \\
\r {22} {\eightit University of Pittsburgh, Pittsburgh, Pennsylvania 15260} \\
\r {23} {\eightit Istituto Nazionale di Fisica Nucleare, University and Scuola
               Normale Superiore of Pisa, I-56100 Pisa, Italy} \\
\r {24} {\eightit Purdue University, West Lafayette, Indiana 47907} \\
\r {25} {\eightit University of Rochester, Rochester, New York 14627} \\
\r {26} {\eightit Rockefeller University, New York, New York 10021} \\
\r {27} {\eightit Rutgers University, Piscataway, New Jersey 08854} \\
\r {28} {\eightit Academia Sinica, Taiwan 11529, Republic of China} \\
\r {29} {\eightit Superconducting Super Collider Laboratory, Dallas,
Texas 75237} \\
\r {30} {\eightit Texas A\&M University, College Station, Texas 77843} \\
\r {31} {\eightit Texas Tech University, Lubbock, Texas 79409} \\
\r {32} {\eightit University of Tsukuba, Tsukuba, Ibaraki 305, Japan} \\
\r {33} {\eightit Tufts University, Medford, Massachusetts 02155} \\
\r {34} {\eightit University of Wisconsin, Madison, Wisconsin 53706} \\
\r {35} {\eightit Yale University, New Haven, Connecticut 06511} \\
\end{center}
\vskip 0.3in
 \renewcommand{\baselinestretch}{1.5}
\begin{abstract}
    This paper presents the first direct measurement of the
$B$ meson differential cross section, $d\sigma/dp_T$, in
$p\overline{p}$ collisions at $\sqrt{s}=1.8$ TeV using a sample of
 $19.3 \pm 0.7$ pb$^{-1}$ accumulated by
the Collider Detector at Fermilab (CDF).
The cross section is measured in the
central rapidity region $|y| < 1$ for $p_T(B) > 6.0$ GeV/$c$ by fully
reconstructing the $B$ meson decays $B^{+}\rightarrow J/\psi K^{+}$ and
$B^{0}\rightarrow J/\psi K^{*0}(892)$, where $J/\psi \rightarrow
\mu^+\mu^-$ and $K^{*0} \rightarrow K^+ \pi^-$. A comparison is made to
the theoretical QCD prediction calculated at next-to-leading order.
\end{abstract}

\vskip 0.3in
 \renewcommand{\baselinestretch}{1.5}
   The prediction of QCD theory was in
agreement with the first $b$ quark cross section measurements in
$p\overline{p}$ collisions which were made by UA1 at \roots = 630 GeV
\cite{ua1b}.
However, measurements
at \roots = 1.8 TeV
made more recently at the Collider Detector at Fermilab (CDF)
\cite{cdfinc,hugh,psinc}
have cast doubt on whether QCD correctly predicts either the absolute rate
or the shape of the  transverse momentum ($p_T$) distribution.
We present the first direct measurement of the $B$ meson differential
cross section, $d\sigma/dp_T$, in hadronic collisions by
measuring the mass and momentum of the $B$ mesons decaying into exclusive
final states. This result is a more precise test of the QCD prediction
than those measurements that rely on model-dependent procedures
to infer the $b$ quark content and momentum distribution from inclusive
samples.  The data sample used represents $19.3 \pm 0.7$ pb$^{-1}$
collected by CDF during the 1992-93 run.
$B$ mesons are re\-con\-structed via the de\-cays \mbox{\psik} ~and
\mbox{\pstar(892)}, with $J/\psi$ $\rightarrow$ $\mu^+\mu^-$
and $K^{*0}$ $\rightarrow$ $K^+\pi^-$ and their charge con\-jug\-ates.

     Detailed descriptions of the CDF detector have been provided
\mbox{elsewhere \cite{det}}.  The components relevant to this analysis
are briefly described here.  The $z$-axis of the detector coordinate
system is along the beam direction.
The Central Tracking Chamber
(CTC) is a drift chamber in a 1.4T axial magnetic field, consisting of
nine superlayers, four of which give stereo information.  A particle must have
pseudorapidity $|\eta| < 1$ to pass through all nine superlayers, which
defines the rapidity range used for the cross section measurement.

     A Silicon Vertex Detector (SVX) provides high-resolution
$r$-$\phi$ tracking information near the interaction region \cite{svxnim}.
The SVX detector is 51 cm long and consists of
four layers of silicon microstrip detectors with an innermost radius of 2.9 cm.
Pattern recognition is done by extrapolating tracks from the CTC.
The longitudinal distribution of the primary vertices
is such that SVX information
is available for about 60$\%$ of all tracks. If SVX information
is added to a track, the momentum resolution, $\delta p_T/p_T$, improves
from $\sim$0.002$p_T$ to
$[(0.0009p_T)^2+(0.0066)^2]^{1/2}$.  Surrounding the CTC are electromagnetic
and hadronic calorimeters,
outside of which are the central muon chambers,
segmented into 72 modules which provide about $85\%$ coverage in azimuth in the
pseudorapidity range $|\eta| < 0.6$.

     The selection of $B$ candidates begins by identifying
$J/\psi$ candidates that decay to two muons.
There are three levels of trigger requirements that
must be satisfied for a muon pair to be included in the $J/\psi$ data sample.
At the first trigger level, the muons must have been detected in
the central muon chambers and pass a minimal transverse momentum requirement of
$\sim$1.4 GeV/$c$.
Prompt muons with lower momenta range out in the calorimeters.
The efficiency for this trigger is 90$\%$ at $p_T = 3.8$ GeV/$c$. At the second
trigger level, at least one of the muon-chamber tracks must match a track
found in the CTC
by a hardware track processor.  The efficiency for the track processor
rises from 10$\%$ at 2.3 GeV/$c$ to 90$\%$ at 3.4 GeV/$c$.  At the
third(software)
trigger level, detailed CTC pattern recognition and tracking are done, and
the dimuon invariant mass is required to be within 300 MeV/$c$$^2$ of
the $J/\psi$ mass.

     To improve the purity of the $J/\psi$ sample,  the CTC track is
required to match its associated muon chamber track
to within $3\sigma$ in $r$-$\phi$ and $3.5\sigma$ in $z$.
To match the trigger thresholds, each muon is required to have
$p_T \ge 1.8$ GeV/$c$, and at least one muon is required to have $p_T \ge 2.8$
GeV/$c$.
The fitted number of
$J/\psi$ reconstructed after all the requirements have been imposed is about
45,000.
The dimuon invariant mass is required to be within $4\sigma$ of the known
$J/\psi$ mass \cite{pdg}. The $K^+$ and $K^{*0}$ can\-didates are required to
have
$p_T > 1.25$ GeV/$c$; furthermore, each $K^{*0}$ decay product is required
to have $p_T > 400$ MeV/$c$ to ensure reliable tracking. All pairs of
oppositely charged particle tracks are considered to be candidates for the
$K^{*0}$ decay products.  The $K\pi$ invariant mass is required to be within
50 MeV/$c$$^2$ of the $K^{*0}$ mass.  Unlike combinatoric background,
the invariant mass reconstructed for an actual $B^0$ meson peaks at the $B^0$
mass
when the kaon and pion mass assignments are exchanged.   To avoid double
counting, only the $K^{*0}$
candidate with the mass closest to the $K^{*0}$ mass is used.

   We find the $B$-candidate mass and momentum subject to the constraint that
the decay tracks come from a common vertex and the invariant mass of the dimuon
tracks is equal the $J/\psi$ mass.  We require the confidence level of the fit
to be greater than
0.5$\%$.  The transverse momentum for each $B$ candidate is required to be
greater than 6.0 GeV/$c$.  The proper decay length,
$c\tau \equiv L_{xy}m_B/p_T$, is calculated, where $L_{xy}$ is the
projection of the $B$ vertex displacement
onto the $B$ transverse momentum.  About 75$\%$ of the
background that occurs when a prompt $J/\psi$ is combined with
other tracks from the primary vertex is removed by requiring $c\tau$ to be
greater than 100~$\mu m$.

\begin{figure}
\epsfverbosetrue
\epsfysize=6.0in
\epsfxsize=6.0in
\epsfbox{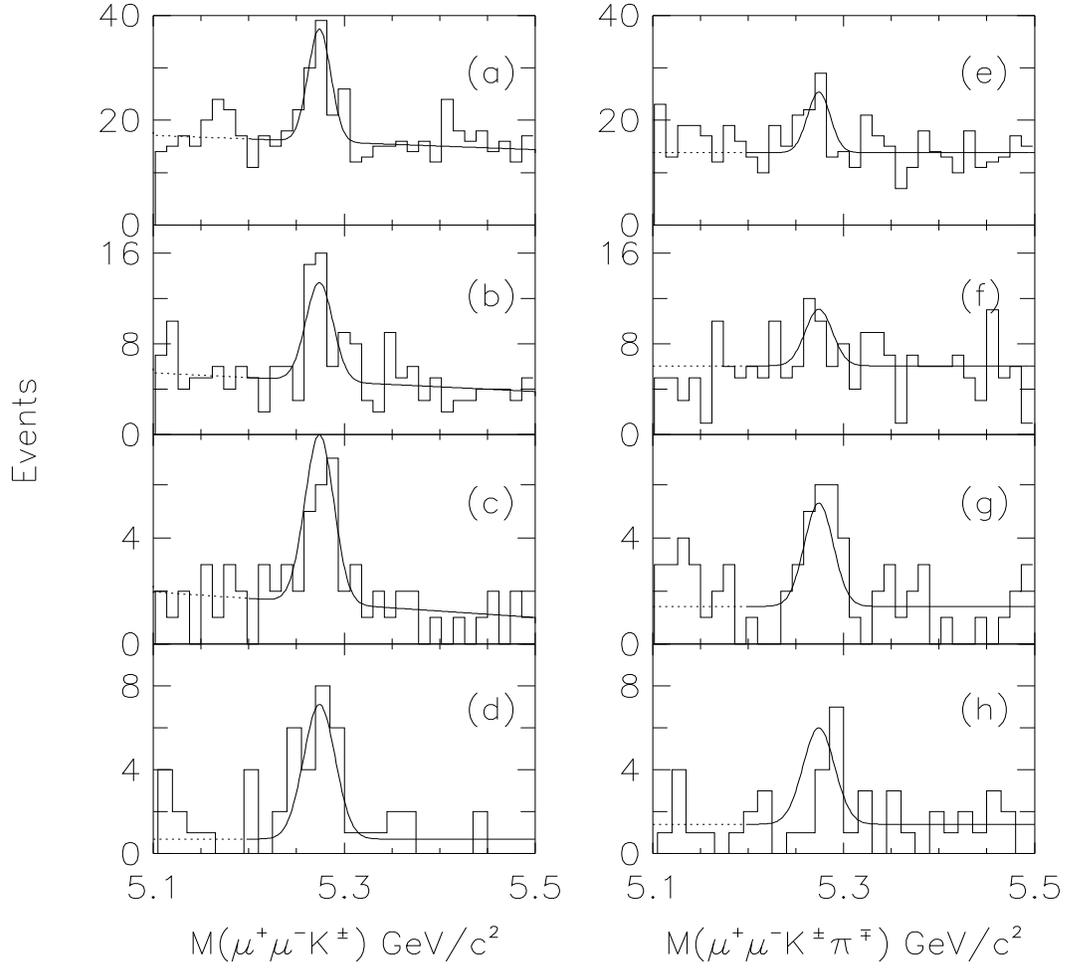}
\vspace*{3pt}
\caption{\label{figg1} $B^\pm$ and $B^0$ meson invariant mass distributions
for the momentum ranges (a,e) 6-9 GeV/$c$, (b,f) 9-12 GeV/$c$,
(c,g) 12-15 GeV/$c$, and (d,h) $>$15 GeV/$c$.}
\end{figure}

    The $B$ candidates are divided
into subsamples in $p_T$ ranges 6-9, 9-12, 12-15, and $>15$ GeV/$c$.
Isospin symmetry in fragmentation, {\em i.e.} equal production of $B^+$ and
$B^0$, is expected to hold for central production \cite{dunietz}.  To make the
best measurement of $d\sigma/dp_T$, the $B^\pm$ and $B^0$ data samples are
combined.
The resulting average $B$ meson cross section as well as the separate $B^+$
cross section are shown in Table I.  The remainder of the paper describes the
analysis for the average cross section in detail.
For each $p_T$ range, the $B^+$ and $B^0$ invariant mass distributions are
fit simultaneously, using an unbinned likelihood method. The relative numbers
of signal events for each $p_T$ bin are constrained to be proportional to
the relative reconstruction efficiencies, $\epsilon(B^0)/\epsilon(B^+)$,
which are shown in Table \ref{tabb1}.
The mass is set to 5.274 GeV/$c$, which is
the value found from a fit to the entire data sample.
The mass range below 5.2 GeV/$c$$^2$ is excluded from the fits since it can
include contributions from higher multiplicity $B$ decay modes.
The slope of the background is constrained to that found from a sample of
candidate events that fail the vertex $\chi^2$ confidence level requirement.
The $B^+$ and $B^0$ invariant mass
distributions for each momentum range are shown in Fig.
\ref{figg1}, and the fitted numbers of events and the statistical uncertainties
are given in Table I.  The systematic uncertainty in the fitting method is
determined to be 9.2$\%$ by taking into account the uncertainties in the
mass, background slopes, and relative reconstruction efficiencies.

   The $B$-meson differential cross section is calculated from the equation,
\begin{equation}
\frac{d\sigma(B)}{dp_T}=\frac{N}{{2\cdot\cal{L}}\cdot A \cdot \epsilon
\cdot F \cdot \Delta p_T}
\end{equation} where $N$ is the number of events observed, $\cal{L}$ is the
integrated luminosity, $A$ is the detector acceptance (including the efficiency
of the kinematic cuts), $\epsilon$ is the
combined tracking and track-matching efficiency,
$F$ is the  branching fraction, and $\Delta p_T$ is the width of the
$p_T$ bin.  The factor of $1/2$ is included because
decays involving both $B$ and $\overline{B}$ mesons have been reconstructed,
but the quoted cross sections are for $B$ mesons only.

     A Monte Carlo simulation employing the next-to-leading order QCD
calculation\cite{nde} with renormalization scale
$\mu_0=\sqrt{{m_b}^2+{p_T}^2}$, where the $b$-quark mass, $m_b$, is equal to
4.75 GeV/$c$$^2$; the MRS$D_0$ proton structure
functions\cite{sti};
the Peterson parameterization\cite{peter} for fragmentation, using a value of
the fragmentation parameter of 0.006; and
a detector and $J/\psi$ trigger simulation was
used to determine the acceptance, shown in Table I.

     Online (third trigger level) and offline tracking efficiencies are
$97 \pm 2\%$ and $98.5 \pm 1.4\%$, respectively.  The efficiency
of the matching requirement between the CTC track and the
muon chamber track segment
is $98.7 \pm 1\%$.  Product branching fractions of
$(6.55 \pm 1.01)\times 10^{-5}$ and
$(6.67 \pm 1.21)\times 10^{-5}$ \cite{cleo} were used for the \mbox{\psik} and
\mbox{\pstar} decays, which include the $J/\psi \rightarrow
\mu^+\mu^-$ and $K^{*0} \rightarrow K^+ \pi^-$ branching ratios.  These
branching ratios are included in the relative reconstruction efficiencies given
in Table \ref{tabb1}.

    Varying the $b$-quark mass and the QCD renormalization scale
used in the Monte Carlo simulation within their uncertainties changes
the calculated acceptance by $\pm$2$\%$ \cite{thesis}. The systematic
uncertainty
in the $J/\psi$ efficiency due to the trigger parameterization was determined
to be
$5.5\%$.  Additionally, a systematic uncertainty of $4\%$ is associated
with the reconstruction of kaons that decay inside the CTC volume.  The
efficiency of the 100~$\mu m$ cut on $c\tau$ has an uncertainty of
$4\%$, due to its dependence on the lifetime of
the meson and on the $c\tau$ resolution \cite{thesis}, which
varies from 50 to 300 $\mu m$, depending on whether or not SVX information was
available for the vertex fit.    The $\chi^2$ requirement on the common-vertex
constraint
has an additional 1$\%$ uncertainty, which was determined from the fraction of
$J/\psi$ that fail such a requirement.  Varying the polarization of the
\pstar~ decay
products in the range (80$\pm$10)$\%$\cite{cleo2} changes the calculated
\noindent acceptance by $\pm5.7\%$.      Combining these values in quadrature,
the
reconstruction efficiency has overall systematic uncertainties of $11.1\%$
for the $B^+$ decay and $13.5\%$ for the $B^0$ decay.

\begin{figure}
\epsfverbosetrue
\epsfysize=6.0in
\epsfxsize=6.0in
\epsfbox{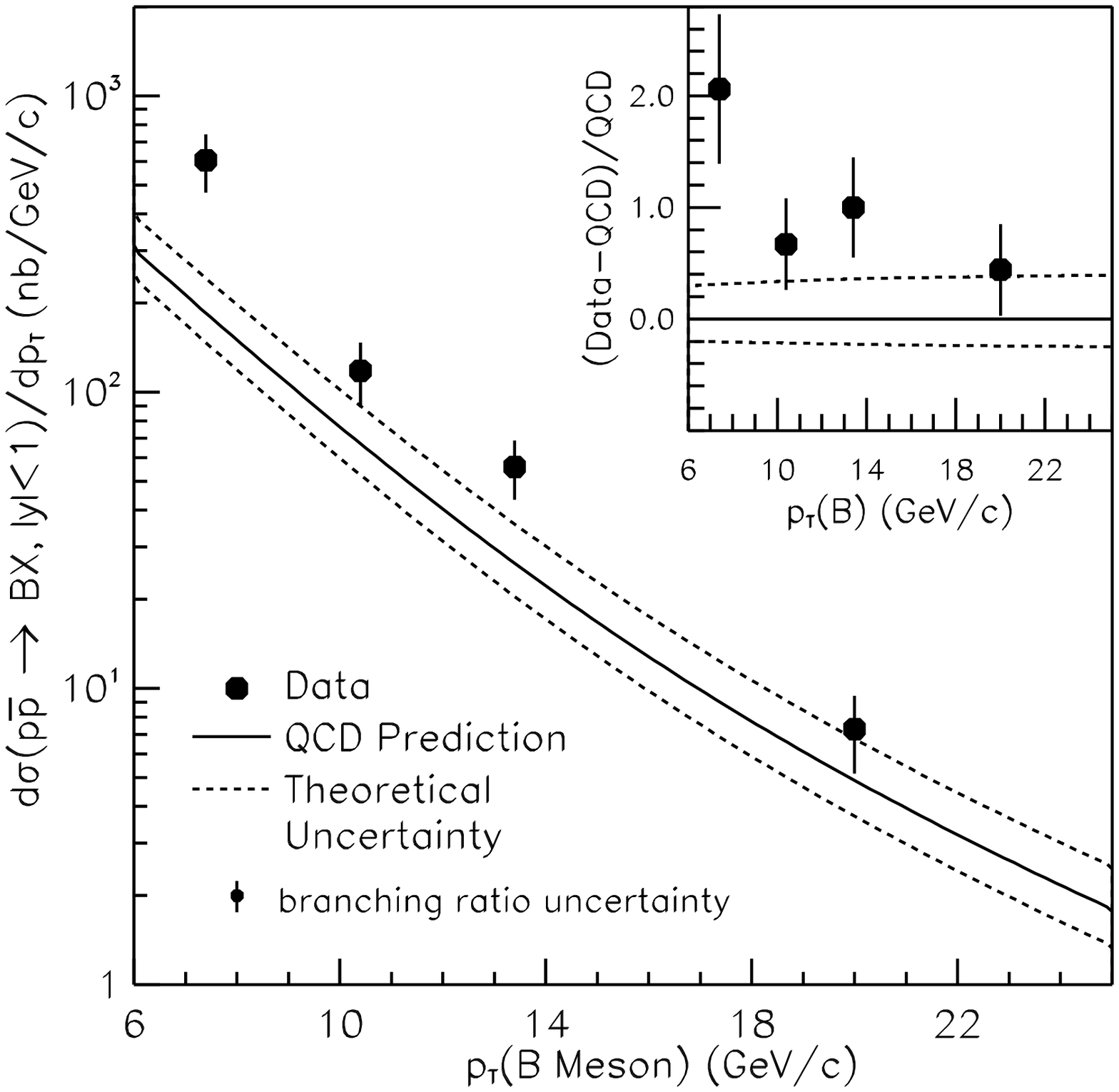}
\vspace*{3pt}
\caption{\label{figg3} Average $B$ meson differential cross section compared
to the
QCD prediction.  The branching ratio uncertainty, which contributes 11.9$\%$ to
the cross section uncertainty, is shown separately.  The dashed curves indicate
the uncertainty in the theoretical prediction.  The inset plot shows
\mbox{(data-QCD)/QCD} on a linear scale.}
\end{figure}

   The cross section results are listed in Table \ref{tabb1} and plotted in
Fig. \ref{figg3}, where the common
branching ratio uncertainty of 11.9$\%$ (included in Table I) is
shown separately.
The differential cross section measurement at 20 GeV/$c$
is obtained by dividing the integrated cross section above 15
GeV/$c$ by an effective bin width determined from the data to be $12.7 \pm 1.3$
GeV/$c$.  The point is plotted at the mean $p_T$ value of the data points.
The integrated cross-section above 6 GeV/$c$ is $2.39\pm0.32\pm0.44$ $\mu$b.
The solid curve in Fig. \ref{figg3} shows the $B$ meson differential cross
section
predicted by the QCD-based Monte Carlo, while the dashed curves indicate the
variation
associated with uncertainty in the $b$ quark mass, the fragmentation parameter,
and the renormalization scale \cite{MC}.  The curves include the generally used
assumption that $75\%$
 of $\overline{b}$ quarks fragment in equal amounts to $B^+$ and
$B^0$ mesons \cite{hugh,l3}.
The visual comparison between data and theory is aided by plotting
\mbox{(data-QCD)/QCD} on a linear scale, as shown in the inset to
Fig. \ref{figg3}.  To determine the level of agreement between the data and the
theoretical prediction, the predicted cross section is fitted to the
measurements, holding the shape constant and varying the magnitude.  The
fit yields an overall scale factor
of $1.9 \pm 0.2 \pm 0.2$, with a confidence level of $20\%$.  In conclusion,
we find that the shape of the $B$ meson differential cross section presented
here is adequately described by next-to-leading order QCD, while the
absolute rate is at the limits of that predicted by typical variations
in the theoretical parameters.  It will be interesting to compare these
data to higher order calculations as they become available.

     We thank the Fermilab staff and the technical staffs of the
participating institutions for their vital contributions.  This work was
supported by the U.S. Department of Energy and National Science Foundation;
the Italian Istituto Nazionale di Fisica Nucleare; the Ministry of Education,
Science and Culture of Japan; the Natural Sciences and Engineering Research
Council of Canada; the National Science Council of the Republic of China;
the A. P. Sloan Foundation; and the Alexander von Humboldt-Stiftung.

\r {(a)} {\eightit Visitor.}

\newpage

 \renewcommand{\baselinestretch}{1.0}
\begin{table}
\caption{\label{tabb1}\mbox{$B$ meson differential cross section,}
\mbox{$d\sigma(|y|<1.0)/dp_T$} (nb/GeV/$c$).
The uncertainty listed for the relative
reconstruction efficiency, $\epsilon(B^0)/\epsilon(B^+)$, includes the
branching ratio uncertainty.  The uncertainties shown for the number of
events are statistical only.}
\begin{tabular}{ccccccc}
 $\left<p_T\right>$ & Acceptance & $\epsilon(B^0)/$ &
No. of & $B^+$ & Total No. &  Average $B$\\
 GeV/$c$ & for $B^+$ $\%$ & $\epsilon(B^+)$  &  $B^\pm$ Events & Cross section
&  of Events & Cross section \\
\hline
 7.4 & 1.28 $\pm$ 0.01 & $0.52 \pm 0.12$ & 51 $\pm$ 10 &
{$ 644 \pm 126 \pm 139 $} &72 $\pm$ 12 &
{$ 596 \pm 103 \pm 106 $} \\
 10.4 & 3.71 $\pm$ 0.04 & $0.58 \pm 0.14$ &31 $\pm$ 7 &
{$ 134 \pm 30 \pm 29 $} &42 $\pm$ 9 &
{$ 116 \pm 24 \pm 21 $} \\
 13.4 & 6.45 $\pm$ 0.06 & $0.61 \pm 0.15$ &20 $\pm$ 5 &
{$ 50 \pm 13 \pm 11 $} &35 $\pm$ 7 &
{$ 55 \pm 10 \pm 10 $} \\
 20.0 & 9.55 $\pm$ 0.10 & $0.71 \pm 0.17$ &24 $\pm$ 4 &
{$ 9.5 \pm 1.6 \pm 2.6 $} & 31 $\pm$ 6 &
{$ 7.2 \pm 1.4 \pm 1.8 $} \\
\end{tabular}

\end{table}

\begin{thebibliography}{abcd}
\bibitem{ua1b} UA1 Collaboration, C. Albajar {\em et al.},
Phys. Lett. B {\bf 186}, 237 (1987);
{\bf 213}, 405 (1988);
{\bf 256}, 121 (1991).

\bibitem{cdfinc} CDF Collaboration, F. Abe {\em et al.}, Phys. Rev. Lett.
{\bf 71}, 500 (1993); {\bf 71}, 2396 (1993);
Report No. Fermilab-Pub-94/131-E (to be published).

\bibitem{hugh} CDF Collaboration, F. Abe {\em et al.}, Phys. Rev. Lett.
{\bf 68}, 3403 (1992); Phys. Rev. D {\bf 50}, 4252 (1994).


\bibitem{psinc} CDF Collaboration, F. Abe {\em et al.}, Phys. Rev. Lett.
{\bf 69}, 3704 (1992); {\bf 71}, 2537 (1993). These papers assumed no direct
production of $J/\psi$ or $\psi(2S)$.

\bibitem{det} CDF Collaboration, F. Abe {\em et al.}, Nucl. Instrum. Methods
Phys. Res., Sect. A {\bf 271}, 387 (1988).

\bibitem{svxnim} D. Amidei {\em et al.}, Nucl. Instrum. Methods Phys.
Res., Sect. A {\bf 350}, 73 (1994).

\bibitem{pdg} M. Aguilar-Benitez {\em et al.}, Phys. Rev. D {\bf 50}, 1173
(1994)

\bibitem{dunietz} I. Dunietz and J. Rosner, Report No. Fermilab-Pub-94/298-T.

\bibitem{nde} P. Dawson {\em et al.}, Nucl. Phys.
{\bf B327}, 49 (1988); M. Mangano {\em et al.}, Nucl. Phys.
{\bf B373}, 295 (1992).

\bibitem{sti} A. Martin, R. Roberts and J. Stirling, Report No. RAL-92-021
(to be published).

\bibitem{peter} C. Peterson {\em et al.}, Phys. Rev. D {\bf 27}, 105 (1983);
J. Chrin, Z. Phys. C {\bf 36}, 163 (1987).

\bibitem{cleo}
CLEO Collaboration, M.S. Alam {\em et al.},
\newblock Phys. Rev. D {\bf 50}, 43 (1994).

\bibitem{thesis} M. W. Bailey, Ph.D. thesis, Purdue University, 1994.

\bibitem{cleo2}
CLEO Collaboration, M.S. Alam {\em et al.},
\newblock Phys. Rev. D {\bf 50}, 43 (1994); ARGUS Collaboration, H. Albrecht
{\em et al.}, Phys. Lett. B {\bf 340}, 217 (1994); CDF Collaboration,
F. Abe {\em et al.}, Report No. Fermilab-Conf-94/216-E.

\bibitem{MC} The $b$ quark mass is varied between 4.5 and 5.0 GeV/$c$$^2$,
the renormalization scale is varied between $\mu_0/2$
and $2\mu_0$, and the fragmentation parameter is varied between 0.004 and
0.008.

\bibitem{l3}  L3 Collaboration, B. Adeva {\em et al.}, Phys. Lett. B
{\bf 252}, 703 (1990); ALEPH Collaboration, D. Decamp {\em et al.}, Phys. Lett.
B {\bf 258}, 236 (1991); UA1 Collaboration, H. C. Albajar {\em et al.}, Phys.
Lett. B {\bf 262}, 171 (1991); CDF Collaboration, F. Abe {\em et al.}, Phys.
Rev. Lett. {\bf 67}, 3351 (1991).

\end{thebibliography}
\end{document}